\documentclass[prx,twocolumn,amsmath,amssymb,superscriptaddress]{revtex4-1}
\usepackage{graphicx,,amssymb,amsmath,subfigure}
\usepackage{bm}
\usepackage{dcolumn}
\usepackage{subfigure}
\usepackage[usenames]{color}

\newcommand{\para}[1]{\left(#1\right)}

\newcommand{\sd}[0]{\ \ \ }

\begin{document}
\title{Topological edge states at a tilt boundary in gated multi-layer graphene}
\author{Abolhassan Vaezi}
\affiliation{Cornell University, Ithaca, NY}
\author{Yufeng Liang}
\affiliation{Washington University, St.Louis, MO}
\author{Darryl H. Ngai} 
\affiliation{Cornell University, Ithaca, NY}
\author{Li Yang} 
\affiliation{Washington University, St.Louis, MO}
\author{Eun-Ah Kim}
\affiliation{Cornell University, Ithaca, NY}

\begin{abstract}
Despite much interest in engineering new topological surface(edge) states using structural defects, such topological surface states have not been observed yet. We
show that recently imaged 
tilt boundaries in gated multi-layer graphene 
should support 
topologically protected gapless edge states. We approach the problem from two perspectives:  the microscopic perspective of a tight-binding model and an ab-initio calculation on a bilayer, and  the symmetry protected topological (SPT) states perspective for a general multi-layer.   
Hence we establish the tilt boundary edge states as the first concrete example of edge states of symmetry enriched ${\mathbb Z}$-type SPT, protected by no valley mixing, electron number conservation, and time reversal $T$ symmetries.  
Further we discuss possible phase transitions between distinct SPT's upon symmetry changes. Combined with recently imaged tilt boundary network, our findings offer a natural explanation for the long standing puzzle of sub-gap conductance in gated bilayer graphene, which can be tested through future transport experiments on tilt boundaries. In particular, the tilt boundaries offer an opportunity for  in-situ imaging of topological edge transport 
\end{abstract}
\maketitle
\section{Introduction}
Graphene has garnered interest from broad spectrum of communities, 
ranging from those aiming at atomic scale circuit devices to those searching for new topological phases. Both communities sought after ways to gap the massless Dirac spectrum. The realization of a gate-induced band-gap in the Bernal stacked bi-layer graphene~\cite{Oostinga:2008fk} following the prediction in Ref.~\cite{PhysRevB.74.161403} brought the holy grail of graphene based transistor one step closer to reality. 
 However, the sub-gap conductance measured by~\textcite{Oostinga:2008fk} with weak temperature dependence well below the optically measured gap as large as 250 meV\cite{Zhang:2009fk}
introduced a new puzzle and obstacle: the gapped bilayer is not as insulating as it should be. Dominant transport along physical edge of the samples proposed earlier by \textcite{Li:2011fk} have been ruled out by Corbino geometry measurements\cite{doi:10.1021/nl102459t}, which observed two-dimensional variable range hopping type temperature dependence, independent of geometry. In this paper we predict existence of topological gapless channel of transport along recently imaged AB-BA tilt boundary network
\cite{McEuen2012,2012arXiv1207.5427H,Brown2012}
which solves the puzzle.

The predicted topological edge state holds the promise of the first realization of topological surface(edge) state hosted by structural topological defect. Though there has been much theoretical interest in topological gapless modes hosted by structural topological defects\cite{Ran:2009uq,PhysRevX.2.031013,PhysRevB.82.115120} no such topological gapless mode has been observed so far. The lattice dislocations in three dimensional crystals previously discussed occur deep in the sample that is not directly accessible. However, the tilt boundary of interest  have recently been observed\cite{McEuen2012,2012arXiv1207.5427H,Brown2012}. 
The tilt boundary is a structural topological line defect along which each neighboring layer is displaced by one inter-atomic spacing. Such defect can occur due to the third dimension added by the stacking of the graphene layers; it forms a boundary between two inequivalent stacking structures frequently referred to as AB and BA.
 Here we show that the tilt boundaries host gapless modes of topological origin and form the first example of a naked structural defect hosting topological electronic states.  

\begin{figure}[t]
\includegraphics[width=0.5\textwidth]{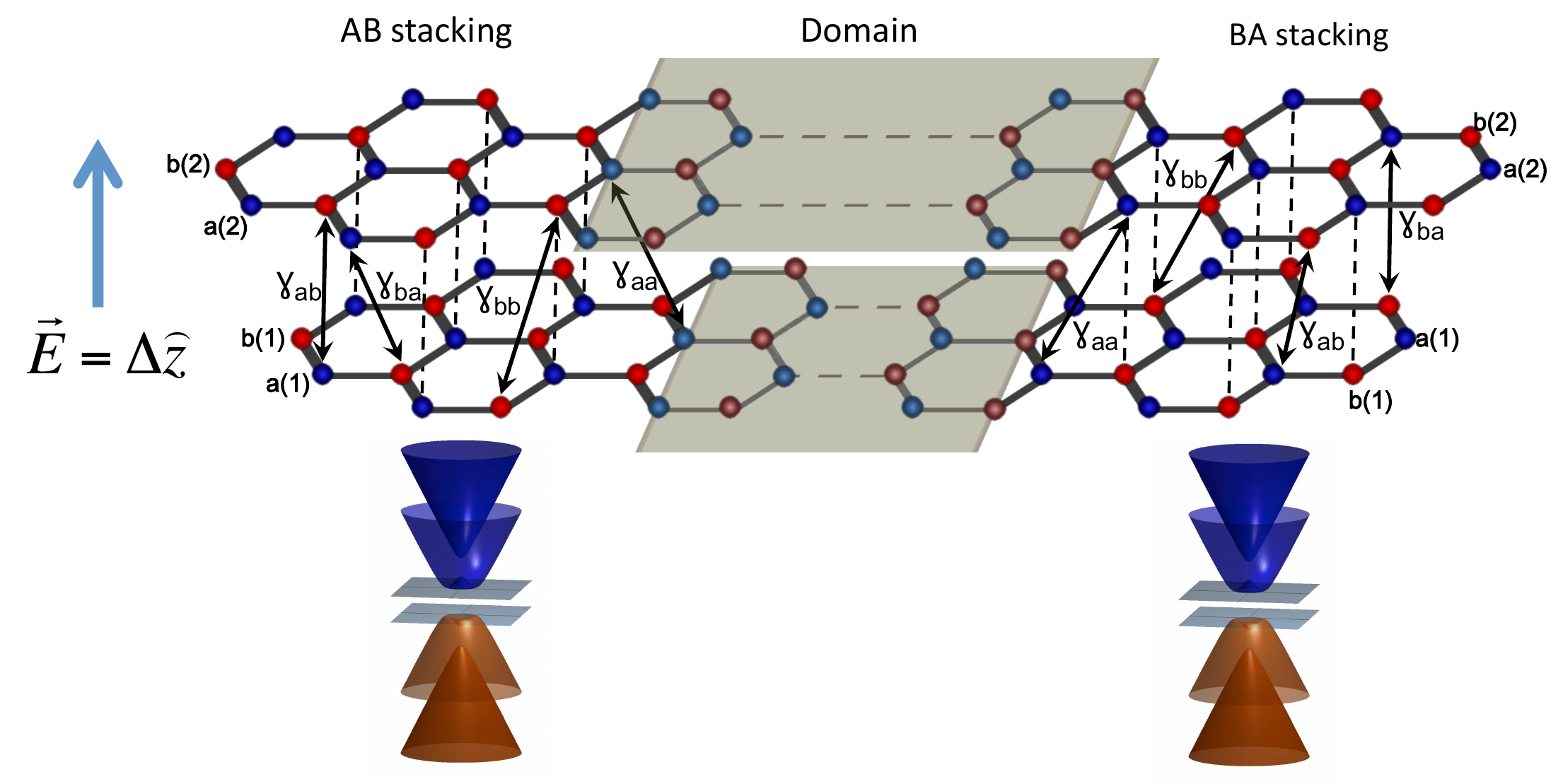}
\caption{A typical AB-BA tilt boundary under strain.
The blue (red) filled circles mark the $a$ ($b$) sublattice sites. $\gamma_{ij}$ represent hopping matrices for a tight-binding model. 
} \label{fig:bilayer}
\end{figure}

Topological aspects of gapped multi-layer graphene have been previously discussed\cite{PhysRevLett.99.236809} and it was pointed out that they should exhibit quantum valley Hall effect with corresponding edge states. However, to this date there has been no experimental detection of proposed edge state~\cite{Li:2011fk,Martin_Topological_2008}. Moreover, little is known about how the topological aspects of gapped multi-layer graphene relates to topological insulators~\cite{PhysRevLett.95.226801,PhysRevLett.96.106802}.  The idea of classifying different topological insulator (superconductor) candidates based on symmetries~\cite{Classification-2} have played a key role in the field of topological insulators. In particular the observation that additional symmetries such as the crystalline symmetries can enlarge the possibilities of topological phases\cite{Classification-3,PhysRevLett.106.106802} led to the discovery of three-dimensional topological crystalline insulators\cite{Dziawa:2012fk}. On one hand we propose feasible experiments to detect topological edge states at naturally occurring tilt boundaries. At the same time, we make first concrete application of the SPT approach~\cite{Classification-3} for two dimensional (2D) system and 
study a large class of gapped graphene systems placing the quantum valley Hall insulator in the larger context and predicting conditions for topological superconductors. 

The rest of the paper is organized as follows. In section \ref{sec:micro} we show that a AB-BA tilt boundary in gated bilayer graphene supports gapless edge states through explicit microscopic calculations. Specifically we consider an abrupt boundary in tight-binding model and then investigate the effect of strain using ab-initio calculation. In section \ref{sec:topo} we show that these edge-states are protected by no valley mixing, electron number conservation, and time reversal ($T$) symmetries within the framework of SPT. Hence we identify chirally stacked gated $N$-layer graphene layers as {\it time-reversal symmetric}  ${\mathbb Z}$-type SPT. In section \ref{sec:expt} we discuss experimental implications. 
Finally in section \ref{sec:summary} we summarize the results and comment on practical implications. 

\section{Bi-layer tilt boundary edge state}
\label{sec:micro}
\begin{figure}[h]
\subfigure[]{
\includegraphics[width=0.4\textwidth]{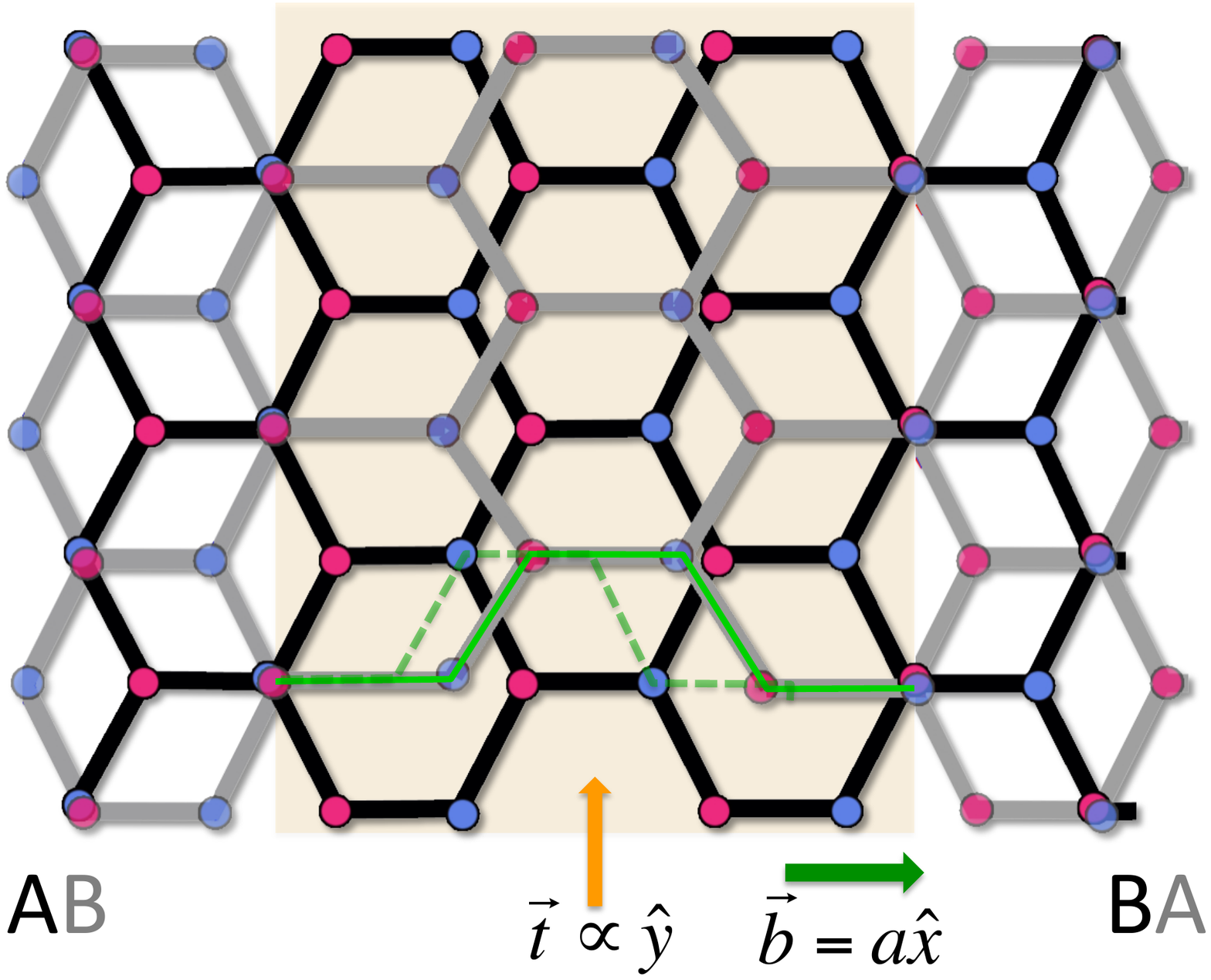}}\\
\begin{tabular}[b]{c}
\subfigure[]{
\includegraphics[width=0.4\textwidth]{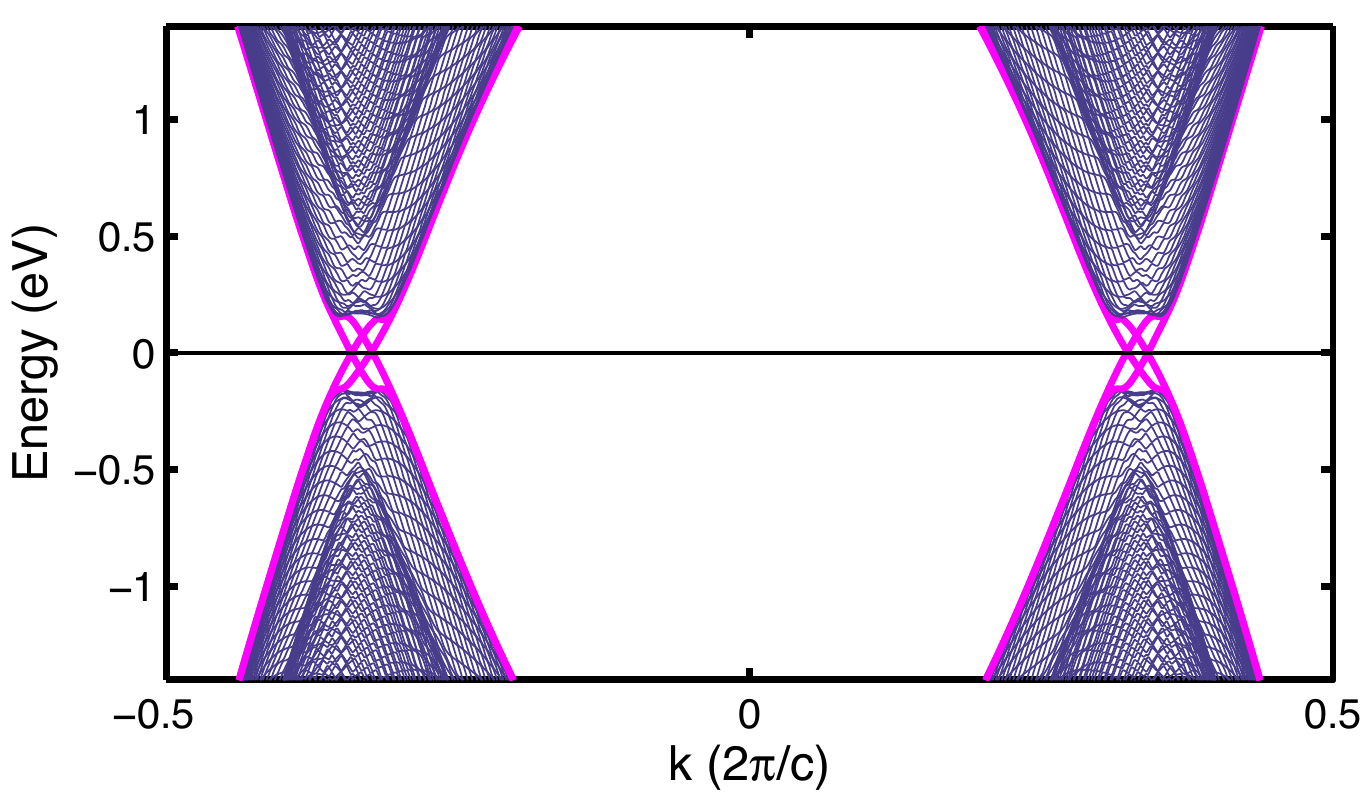}}\\
\subfigure[]{
\includegraphics[width=0.3\textwidth]{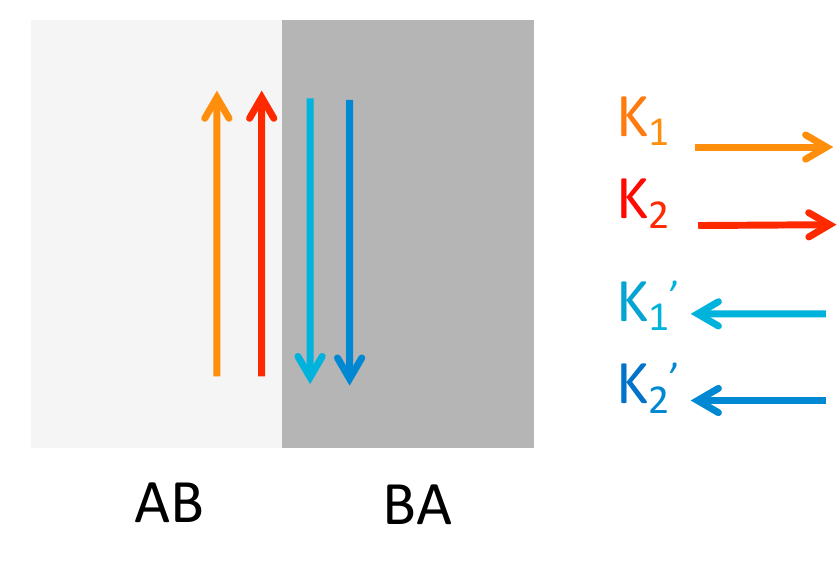}}
\end{tabular}
\caption{A tight binding study in the presence of abrupt AB-BA tilt boundaries in gated bilayer graphene. (a) Schematic representation of the domain wall under strain. 
The grey (black) lines denote the upper (lower) layer, and blue (red) solid circles denote the A (B) sublattice points. 
The tilt boundary along tangent vector $\vec{t}\parallel\hat{y} $ is shaded. The Burger's vector $\vec{b}=a\hat{x}$ (green arrow) for interatomic spacing $a$ accounts for the difference between the green solid line and the green dashed line.
 (b) The resulting band structure. Edge states are marked in magenta. (c) Schematics of valley-momentum locked edge states at a tilt boundary.}
\label{fig:domainwall}
\end{figure}
Fig.~\ref{fig:bilayer} and Fig.~\ref{fig:domainwall}(a) show tilt boundaries of interest in gapped Bernal stacked bi-layer graphene.  In the case sketched, strain is concentrated at the tilt boundary with the top layer stretched by one inter-atomic spacing with respect to the bottom layer. 
For a general orientation,
tilt boundaries can involve both strain and shear. As the tilt boundaries in layered graphene form a type of topological line defects in structure, they can be characterized using the tangent vector $\vec{t}$ and the Burger's vector $\vec{b}$. The tangent vector $\vec{t}$ points along the tilt boundary which can point along any direction with respect to the 
Burger's vector $\vec{b}$. 
When the tilt boundary only involves strain as in the case depicted in Fig.~\ref{fig:bilayer}  and Fig.~\ref{fig:domainwall}(a), the $\vec{b}$ is perpendicular to $\vec{t}$. In the opposite extreme limit of $\vec{b}\parallel \vec{t}$, shear is concentrated at the boundary. Independent of the angle between $\vec{b}$ and $\vec{t}$, the Burger's vector magnitude is the inter-atomic spacing i.e. $|\vec{b}|=a$ for a bilayer system, as it is shown explicitly for the strain tilt boundary in Fig.~\ref{fig:domainwall}(a). Since $|\vec{b}|$ is a fraction of the Bravis lattice primitive vector magnitude $\sqrt{3}a$, 
the bilayer domain boundary is a partial dislocation from quasi two-dimensional view. In a general mult-layer a vertical array of these partial dislocations form a tilt-boundary. 
 In typical samples, the domain wall separating the AB and BA stacked domains has substantial width spanning
5-20 nm and the angle between $\vec{b}$ and $\vec{t}$ ranges between $0^{\rm o}$ and $90^{\rm o}$~\cite{McEuen2012}.

Fig.~\ref{fig:bilayer} allow us to makes two important microscopic  observations about tilt-boundaries. (1) As the boundary requires a shift of one-layer with respect to the other by one inter-atomic spacing along the bond direction, there are three natural directions for the tilt-boundary to run for each fixed angle between $\vec{b}$ and $\vec{t}$. (2) The boundary is arm-chair for $\vec{b}\parallel\vec{t}$ (pure shear) whereas it is zigzag for $\vec{b}\perp\vec{t}$ (pure strain case shown in Fig.~\ref{fig:bilayer}). Based on observation (1) we expect a given type of tilt boundary to possibly form a triangular network seen in experiments\cite{McEuen2012,2012arXiv1207.5427H,Brown2012}. The observation (2) combined with earlier microscopic studies of boundary condition effects on edge states in Ref.~\cite{DW-7} implies that electronic spectrum at tilt-boundaries with $\vec{b}\parallel\vec{t}$ will be gapped though the gap magnitude will be small when the tilt boundary is spread over finite width. 

In this section, we consider the electronic structure of tilt boundaries  with $\vec{b}\perp\vec{t}$ and return to more general case in the section \ref{sec:expt}. As it is shown schematically in Fig.~\ref{fig:bilayer}, the bulk of each domain is gapped in the presence of inter-layer hopping and the external electric field. The latter is important for breaking the inversion symmetry between the layers and gapping otherwise touching bands~\cite{PhysRevB.74.161403,Oostinga:2008fk}.
Below we present two separate microscopic calculations of the AB-BA tilt boundary electronic structure for $\vec{b}\perp\vec{t}$, which shows gapless edge states.

\subsection{Tight-binding model}

We consider a tight-binding Hamiltonian with nearest-neighbor
intra- and inter-layer hopping. For the AB-stacking region (see Fig.~\ref{fig:domainwall}(a)), 
\begin{eqnarray}\label{tbh}
H_{AB} =&& -t \sum\limits_{i=1}^{2} \sum\limits_{m,n} {a^{(i)}_{m,n}}^\dagger \left(b^{(i)}_{m,n}+b^{(i)}_{m-1,n}+b^{(i)}_{m,n-1}\right)\cr
&& + \Delta\sum\limits_{i=1}^{2}\sum\limits_{m,n}(-1)^i  \left({a^{(i)}_{m,n}}^\dagger a^{(i)}_{m,n} + {b^{(i)}_{m,n}}^\dagger b^{(i)}_{m,n} \right)\cr
 &&- t_{\perp} \sum\limits_{m,n} {a^{(1)}_{m,n}}^\dagger b^{(2)}_{m,n} + h.c., 
\end{eqnarray}
where $i=1,2$ is a layer index,  $t$ and $t_{\perp}$ are intra-layer and interlayer nearest neighbor hopping respectively, and $\Delta$ is the chemical potential difference between two layers due to the gate voltage. (m,n) labels the position of the two site unit-cell with $a^{(i)}_{m,n}$ and $b^{(i)}_{m,n}$ annihilating electrons at the two sites of layer $i$.  For the BA stacked region, the only change is in the inter-layer term with  $(- t_{\perp}
\sum\limits_{m,n} {b^{(1)}_{m,n}}^\dagger a^{(2)}_{m,n} + h.c)$ replacing the inter-layer term in 
$H_{AB}$.
  As we address the effect of strain through ab-initio simulation, we focus here on  a sharp tilt boundary as shown in Fig.~\ref{fig:domainwall}. 

We plot the energy spectrum in  Fig.~\ref{fig:domainwall}(b) with the model parameters set to be $t=2.8$eV, $t_{\perp}=0.4$eV, and $\Delta=0.5$eV. The size of the system was 200 unit-cell in each direction under periodic boundary condition with each domain spanning  100 unit-cell width separated by two sharp tilt boundaries. From the spectra it is clear that 
$K$ and $K'$ valleys each have two edge states per spin. 
Further investigation of the wave function shows right and left moving edge states associated with the given valley are spatially separated between the two edges: the edges offer valley filtering (see Fig.\ref{fig:domainwall}(c)).

\subsection{\emph{Ab Initio} Simulation}
\begin{figure}[b]
\includegraphics[width=0.48\textwidth]{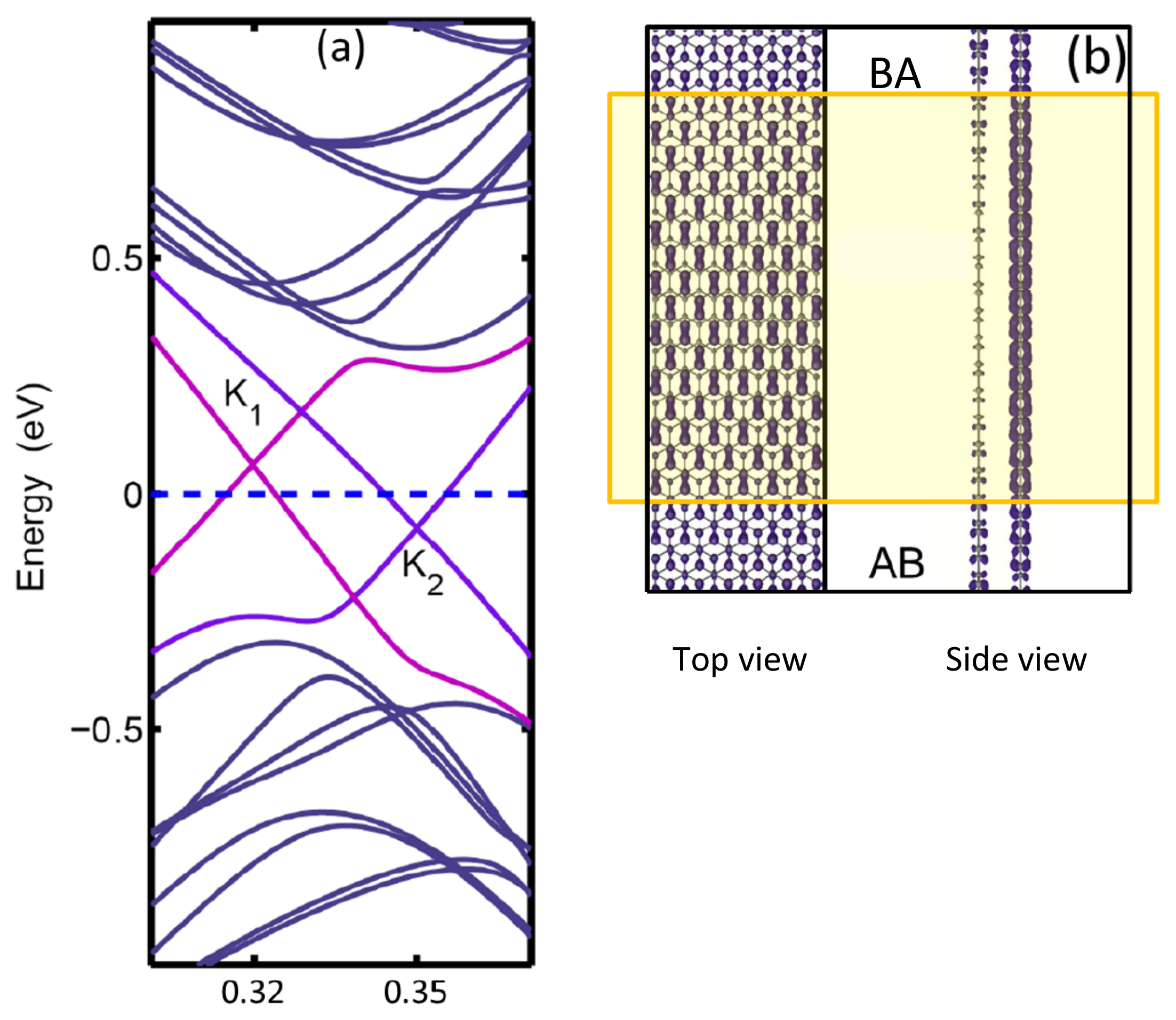}
\caption{(a) The energy bands of AB-BA bilayer graphene connected
two domain walls depicted in Fig.~\ref{fig:domainwall}. The Fermi
level is indicated with the blue dashed line and the bands with
marked linear dispersion relation intersecting near the Fermi
level are outlined by the purple (light) curves.  (b) Top and side
views of the charge distribution for a state near the $K_1$ point
of (a). The yellow rectangles indicate the AB-BA domain
wall structure. } \label{fig:LDABand}
\end{figure}
The electronic structure in realistic tilt boundary will be affected by both the span over which the lattice structure transition from AB to BA stacking, and the strain concentrated at the tilt boundary. To address these issues 
we further carried out
first-principles calculation using density functional theory (DFT)
within the local density approximation (LDA)~\cite{DFT-1,DFT-2}.
We constructed a periodic supercell with two tilt boundaries characterized by Burgers vectors identical to the ones considered within tight binding calculation, but the domain wall is set to have finite width over which the one unit-cell mismatch is spread. We have tried several configurations with domains and tilt boundaries of different widths to find qualitatively similar results. In the rest of this paper we focus on a representative example with 3.1 nm wide domain wall between 1.5 nm wide domains. The choice of narrow width for the domains was due 
to  the limitation in the simulation
capability. To this 2-D system, we then applied a slightly exaggerative perpendicular electric field of 5 V/nm. By relaxing the domain boundaries until the forces reaches below 0.5 eV/nm, we took the effect of both strain and the width into account. We have tried several configurations with domains and tilt boundaries of different width but otherwise similar setting to find qualitatively similar results. 

The DFT-LDA simulation results are presented in Fig.
~\ref{fig:LDABand}, which confirms the existence of the
gapless edge states predicted by the tight-binding model. 
The electronic structure along the extended direction of edge is shown in 
 Fig. ~\ref{fig:LDABand}(a), focusing on the region near $K$ point. 
In this figure two gapless 1D Dirac dispersion is clearly resolved from the gapped bulk specta, with  two distinct Dirac points $K_1$ and $K_2$ in the vicinity of the $K$ point. 
The strain concentrated at the tilt boundary causes
energy splitting of $K_1$ and $K_2$ states; the energy of 1D Dirac points is increased by compressive strain and decreased by tensile strain. This energy splitting will become negligibly small in realistic tilt boundaries with much wider span, as the strain will become smaller. 
 Fig. ~\ref{fig:LDABand}(a) shows that the existence of gapless edge states found in our tight-binding calculation are robust against long range perturbations such as tilt boundary width or subtle bond-length variation inside the domain wall, as well as the strain at the tilt boundary.

We now turn to the spatial distribution of the gapless edge states. 
For illustration, Fig. ~\ref{fig:LDABand} (b) shows the 
 wavefunction amplitude of the state slightly above the $K_1$ point.
As expected from the tight-binding results, the charge distribution is prominent inside the domain
wall but rapidly decays away from the tilt boundary. The Dirac points of edge
states and the decaying feature of edge states always exist,
making us believe it is a universal feature. Meanwhile, the charge
for the states near $K_1$ is highly localized on the layer subject
to compressive strain, regardless of the direction of the applied
electric field. The situation for the states in vicinity of $K_2$
is similar except that the charge prefer to highly localized on
the layer under tensile strain.
Hence forming a layer selective contact to an isolated edge state could be a mechanism for valley filtering\cite{Rycerz:2007fk,PhysRevLett.99.236809}.

Finally, we comment on the so far ignored effect of interaction. 
If the edge boundaries have substantial width, forward scattering part of the Coulomb interaction will be the dominant correlation effect to the edge states and lead to Luttinger liquid behavior
\cite{PhysRevLett.104.216406}. 

\section{Tilt boundary edge states and $\mathbb{Z}$-SPT}
\label{sec:topo}
In order to address the robustness of the edge states, we investigate topological aspects of the low energy effective theory in the continuum limit. We first show that the AB-BA tilt boundary can be mapped to a gate-polarity boundary of uniform bi-layer. Based on this mapping and results of Refs.~\cite{Martin_Topological_2008,DW-7} on the gate-polarity boundary, we discuss the valley Chern number of the tilt boundary edge states.
We than apply the notion of SPT~\cite{Classification-3} and identify chirally-stacked multi-layer graphene as a realization of $\mathbb{Z}$-type SPT, protected by time-reversal ($T$), absence of valley mixing, charge conservation symmetries.   
This identification enables us to address effects of symmetry changes: topological quantum phase transitions. There are recent studies of such perturbations for specific cases such as Rashba spin-orbit coupling~\cite{DW-8} and magnetic ordering~\cite{DW-9}. Through our first application of  SPT classification scheme by \textcite{Classification-3} to a concrete physical system of multi-layer graphene, we obtain an exhaustive systematic study of topological quantum phase transition possibilities.  

\subsection{Valley Chern Number}
The low energy effective Hamiltonian near the $K$ valley for 
uniformly AB or BA stacked bi-layer is
\begin{align}
H_K^{AB/BA}& =  v_{\rm F}k_x (\mu_x
\otimes \lambda_0) + v_{\rm F}k_y (\mu_y \otimes \lambda_0)\label{eq:H-linear}\\
-(&\Delta/2) (\mu_0 \otimes \lambda_z)  +\frac{t_{\perp}}{2} (\mu_x
\otimes \lambda_x) \mp\frac{t_{\perp}}{2} (\mu_y \otimes \lambda_y),\nonumber
\end{align}
where $-$ or $+$ sign should be used for AB or BA stacking respectively. 
In Eq.~\eqref{eq:H-linear},
$v_{\rm F}=3ta/2$, $\mu_{i}$'s are Pauli matrices acting on the sub-lattice indices, and $\lambda_{i}$'s are Pauli matrices acting on the layer indices.  The effective Hamiltonian near $K'$ is $H_{K'}\para{k_x,k_y}=H_{K}\para{-k_x,k_y}$. 
Now it is straight forward to show that BA stacking is equivalent to AB stacking subject to the opposite gate polarity.
At zero field the Hamiltonian of the AB stacked bilayer can be transformed to that of the BA stacking by interchanging the two layers via the following unitary transformation: $H\rightarrow S^\dagger H S$, with $S=\mu_0\otimes\lambda_x$.
For a gated bi-layer however, the gate polarity has to flip since 
\begin{equation}
\mu_0\otimes\lambda_xH_K^{AB}(\Delta)\mu_0\otimes\lambda_x = H_K^{BA}(-\Delta).
\label{eq:equiv}
\end{equation}
Hence at the level of low energy effective theory, the tilt boundary between AB and BA stacking under uniform external field is equivalent to the gate polarity domain wall of structurally uniform bi-layer proposed by \textcite{Martin_Topological_2008}. 

The above equivalence combined with earlier results on valley Chern number of gated chirally stacked multi-layer offers the topological origin of the helical edge states observed in the microscopic calculation of section~\ref{sec:micro}. First for bi-layer and then for general $N$-layers, it was shown that low energy effective theory of chiral stacked $N$-layer under uniform vertical electric field $\Delta$ has finite Chern number per spin for each valley of equal magnitude and opposite sign~\cite{Martin_Topological_2008,DW-6,DW-7,DW-11}: 
\begin{equation}
C_K= -C_K' = \frac{N}{2} {\rm sgn}(t_\perp\Delta).
\label{eq:CK}
\end{equation}
The Chern numbers in Eq.~\eqref{eq:CK} can be obtained by integrating the Berry curvature over momenta $(k_x,k_y)$ continuing the linearized dispersion to infinity. This combined with the equivalence relation of Eq.~\eqref{eq:equiv} means the valley Chern numbers change sign at the tilt boundary. Such sign change leads to  $\Delta C_K=-\Delta C_K=N {\rm sgn} (t_\perp \Delta)$ across the tilt boundary and $N$ branches of valley helical edge states~\cite{DW-7}, as long as the two valleys $K$ and $K'$ remain distinct. Hence the two valley helical edge states per spin observed in section \ref{sec:micro} originate from the valley Chern number change across the tilt boundary, as in the gate polarity boundary edge states~\cite{Martin_Topological_2008,DW-7}. Hence, our prediction is the tilt boundaries will be the first experimentally observed crystalline topological defects to host topological gapless mode due to change in the Chern number.  

\subsection{Chiral Multi-layer as ${\mathbb Z}$-type SPT}
We now apply the procedure for identifying the class of SPT based on symmetries of free fermion Hamiltonian developed by \textcite{Classification-3}, which predicts possible number of protected edge (surface) states. This procedure allows us to consider additional symmetries in the multi-layer graphene in addition to the $C$, $T$, and $P$ taken into account in the pioneering work by  \textcite{Classification-2}, and by \textcite{Classification-1}.
The procedure consists of three steps: (1) Find a gapless Dirac Hamiltonian (by keeping the kinetic term only) with the same symmetries. Then we find all the symmetry preserving mass terms that can gap out the gapless part and are amenable to classification using Clifford algebra. This is based on the assumption that the SPT order is robust as long as the energy gap stays finite and the symmetries remain the same and hence  any gapped Hamiltonian can be adiabatically transformed into a gapped Dirac Hamiltonian. (2) Express the Hamiltonian and the conserved quantities associated with symmetries in the Majorana basis. This leads to the Clifford algebra (i.e. real representation of the Dirac algebra) associated with the gapless part of the Hamiltonian. (3) Find the space of mass matrices that anti-commutes with all the generators of this Clifford algebra. The resulting space may have disconnected pieces, the number of which gives the classification of the SPT. Two mass matrices are topologically distinct if and only if they belong to two different pieces. Applying this procedure to chiral multi-layer graphene will enable us to study phase transitions into different SPTs upon symmetry changes.

For chiral multi-layer graphene system, we assume no valley mixing, electron number conservation $(U(1)_c)$, and time reversal ($T$) symmetries. 
The relevant gapless Dirac Hamiltonian is:
\begin{eqnarray}
&&H=iv_{\rm F}\int d^2x ~\Psi^\dag \para{x}\para{\rho_{1}\partial_{x}+\rho_{2}\partial_{y}}\Psi\para{x},
\end{eqnarray}
where $\Psi^{\rm T}=\para{c_{A,K},c_{A,K'},c_{B,K},c_{B,K'}}$, and $\rho_1=\mu_x\otimes \tau_z\otimes I_{n\times n}$, and $\rho_2=\mu_y\otimes \tau_0 \otimes I_{n\times n}$, in which $n$ is given by the number of layers. This Hamiltonian can be written in the Majorana fermion basis using the following decomposition
\begin{eqnarray}
c_{\mu,\tau}\para{x}=\frac{\gamma_{+,\mu,\tau}\para{x}+i\gamma_{-,\mu,\tau}\para{x}}{2},
\end{eqnarray}
where $\gamma$ denotes the Majorana fermion satisfying
\begin{eqnarray}
 \gamma_{\alpha,\mu,\tau}^2=1\sd,\sd\{\gamma_{+,\mu,\tau},\gamma_{-,\mu,\tau}\}=0,
\end{eqnarray} 
where $\mu$ ($\tau$) denotes the A, or B sublattice ($K$ or $K'$ valley) indices, and $\alpha$ denotes the flavor of the Majorana fermions (+ or -). In the Majorana fermion basis the Hamiltonian is represented as follows:

\begin{eqnarray} \label{MFH}
H=i\int d^2 x ~\eta\para{x}\mathcal{A}\eta\para{x},
\end{eqnarray}
where $\mathcal{A}$ is a real anti-symmetric matrix (differential operator), and $\eta$ is an eight component vector whose components are $\gamma_{\alpha,\mu,\tau}$. 

Now we express the conserved quantities associated with the symmetries of the Hamiltonian in the Majorana fermion basis. First, no valley mixing combined with total electron number conservation symmetry
leads to separate conservation of the electron number at each valley $N_{K}$ and $N_{K'}$. 
Hence the total electron number
$N_c=N_{K}+N_{K'}$, and the valley polarization $N_{V}=N_{K}-N_{K'}$ are conserved. 
In the Majorana fermion basis,
\begin{eqnarray}
&&N_{c,V}=\frac{i}{4}\int d^2x~\eta \para{x}\hat{Q}_{c,V}\eta\para{x}, \quad{\rm with}\cr
&& \hat{Q}_{c}=i\alpha_{y}\otimes \mu_0\otimes \tau_{0} \otimes I, \quad \hat{Q}_{V}=i\alpha_{y}\otimes \mu_0\otimes \tau_{z} \otimes I,~
\end{eqnarray}  
where $\alpha_i$ ($\tau_i$) Pauli matrices act on the Majorana flavors (valley indices). So defined $Q_{c}$ and $Q_{V}$ satisfy $\hat{Q}_c^2=\hat{Q}_{V}^2=-1$. 

Under time reversal symmetry $T$, $K$ and $K'$ valley indices are exchanged i.e., $\hat{T}: c_{\mu,K} \leftrightarrow c_{\mu,K'}$. Hence $T$ acts like $\tau_{x}$ in the valley basis with the matrix part of the time reversal operator satisfying $\hat{T}^2=1$.
 On the other hand, no valley mixing implies the Hamiltonian is invariant under $\para{c_{\mu,K},c_{\mu,K'}}\to \para{c_{\mu,K},-c_{\mu,K'}} $ transformation which acts like $\tau_z$ in the valley basis. Hence, in the presence of no valley mixing symmetry, we can define a new time reversal operator $\hat{\Theta}:    \para{c_{\mu,K}, c_{\mu,K'}} \to \para{ c_{\mu,K'} ,-c_{\mu,K}}$, which acts like $\tau_z\hat{T}=i\tau_y$.
In terms of Majorana fermions
\begin{eqnarray}
&& \hat{\Theta}=\alpha_{0}\otimes \mu_0\otimes \para{i\tau_{y}} \otimes I~~,~~\hat{\Theta}^2=-1.~~
\end{eqnarray}

In order to find the relevant Clifford algebra, we need to form anti-commuting generators in terms of $\mathcal{A}$ in Eq. \eqref{MFH} combined with symmetries, $\hat{Q}_c$, $\hat{Q}_V$, and $\hat{\Theta}$. However, 
the symmetries require  $\left[\mathcal{A},\hat{Q}_{c}\right]=\left[\mathcal{A},\hat{Q}_{V}\right]=0$ and $\left\{\mathcal{A},\hat{\Theta}\right\}$=0. Moreover, symmetry operators satisfy $\hat{Q}_{c}\hat{\Theta}=\hat{\Theta}\hat{Q}_{c}$,  $\hat{Q}_{V}\hat{\Theta}=-\hat{\Theta}\hat{Q}_{V}$, and $\hat{Q}_{c}\hat{Q}_{V}=\hat{Q}_{V}\hat{Q}_{c}$ relations. Using these relations, we find the full set of generators of the relevant Clifford algebra as $\rho_1$, $\rho_2$, and
\begin{eqnarray}
&&\rho_{3}=\hat{\Theta}\hat{Q}_{V}\hat{Q}_{c}\sd,\sd\rho_{4}=\hat{\Theta}\sd,\sd\rho_{5}=\hat{\Theta}\hat{Q}_{V},
\end{eqnarray} 
as $\left\{\mathcal{A},\rho_{i}\right\}=0$ for $i=3,4,5$.  The resulting full set of anti-commutation relations is
\begin{eqnarray}
&&\left\{\rho_{i},\rho_{j}\right\}=2g_{i,j}\sd,\sd  g_{i,j}={\rm diag}\para{1,1,1,-1,-1},
\end{eqnarray}
and it defines a Clifford algebra Cliff(3,2).

Now, we will find the space of mass matrices, $C_M$, that can gap out Dirac Hamiltonian associated with this Clifford algebra in order to obtain SPT classification. The mass term with matrix representation $H_{M}=\frac{i}{4}\sum_{I,J}M_{I,J}\eta_{I}\eta_{J}$  should satisfy the following algebra:
\begin{eqnarray} 
M\rho_{i}=-\rho_{i}M,\quad M^2=-1,
\end{eqnarray} 
where we have normalized $M$. Solving the above equation yields the allowed space for the mass matrix, $C_{M}$. It has been shown \cite{Classification-3}  that $C_M$ which solves Eq. \eqref{C_M} for the case of Cliff$(3,2)$ is
\begin{eqnarray}\label{C_M}
C_{M}=\lim_{n\to \infty} \bigcup_{m=0}^{n} \frac{O(n)}{O(m)\times O(n-m)}.
\end{eqnarray}
The SPT classification is then given by the number of disconnected pieces in the space of mass matrix $C_{M}$ i.e. its zeroth homotopy group: $\pi_{0}\para{C_M}$. Using Eq. \eqref{C_M}, it can be verified that  $\pi_{0}\para{C_M}=\mathbb{Z}$~\cite{Classification-3}. Consequently, each class of the time reversal invariant multilayer graphene in the absence of intervalley scattering is indexed by a $\mathbb{Z}$-valued number: in this case the valley Chern number.  

\begin{table*}
\label{tab1}
  \centering
  \begin{tabular}{| c| c|  c| c| c|}
  \hline
    ${\rm \Theta}$&${\rm No~valley~mixing}$&${U(1)_c}$&${\rm Classification}$&${\rm Examples}$\\
  \hline
    $\checkmark$&$\checkmark$&$\checkmark$&$\mathbb{Z}$&${\rm QVH}$\\
    $\times$&$\checkmark$&$\checkmark$&$\mathbb{Z}\oplus \mathbb{Z}$&${\rm intravalley~QAH}$\\
    $\checkmark$&$\times$&$\checkmark$&${\rm Trivial}$&${\rm trivial~insulator}$\\
    $\times$&$\times$&$\checkmark$&$\mathbb{Z}$&${\rm intervalley~QAH}$\\
\hline
  $\checkmark$&$\checkmark$&$\times$&$\mathbb{Z}$&${\rm TVSC}$\\
    $\times$&$\checkmark$&$\times$&$\mathbb{Z}\oplus \mathbb{Z}$&${\rm intravalley~TSC}$\\
    $\checkmark$&$\times$&$\times$&${\rm Trivial}$&${\rm trivial~superconductor}$\\
      $\times$&$\times$&$\times$&$\mathbb{Z}$&${\rm intervalley~TSC}$\\
\hline
  \end{tabular}
 \caption{Classification of the SPT on multi-layer graphene by considering the presence ($\checkmark$) or absence($\times$) of the time reversal symmetry $\hat{\Theta}\propto i\tau_y$, no valley-mixing and charge conservation $U(1)_c$. Four upper rows classify topological insulator while four lower rows classify topological superconductors.}\label{table:spinless}
\end{table*}

\begin{table*}
\label{tab2}
  \centering
  \begin{tabular}{| c| c|  c| c| c| c| c|c|}
  \hline
    ${\Theta_1~}$&${\Theta_2~}$&${T=\Theta_1\Theta_2}$&${\rm No~valley~mixing}$&${\rm Classification}$&${\rm Examples}$\\
  \hline
  ${\checkmark}$&${\times}$&${\times}$&${\checkmark}$&${\mathbb{Z}}$&${\rm QVH}$\\
  ${\times}$&${\times}$&${\times}$&${\checkmark}$&${\mathbb{Z}\oplus \mathbb{Z}}$&${\rm intravalley~ QAH}$\\
  ${\times}$&${\times}$&${\times}$&${\times}$&${ \mathbb{Z}}$&$~{\rm intervalley~QAH}$\\
 \hline
  ${\checkmark}$&${\checkmark}$&${\checkmark}$&${\checkmark}$&${\mathbb{Z}_2}$&${\rm LAF}$\\
  ${\times}$&${\checkmark}$&${\times}$&${\checkmark}$&${\mathbb{Z}_2\oplus \mathbb{Z}_2}$&${\rm intravalley~topological~insulator }$\\
  ${\times}$&${\checkmark}$&${\times}$&${\times}$&${ \mathbb{Z}_2}$&$~{\rm intervalley~topological~insulator}$\\
  ${\times}$&${\times}$&${\checkmark}$&${\checkmark}$&${\mathbb{Z}}$&${\rm intravalley~QSH}$\\
 \hline
  \end{tabular}
 \caption{Classification of spinful SPT insulators on multi-layer graphene. Three kinds of time reversal operators are considered for classification due to the valley and spin dynamical degrees of freedom: $\hat{\Theta}_1\propto\tau_x$ exchanges two valleys, $\hat{\Theta}_2\propto i\sigma_y$ which acts on the spin indices flips spin, and $\hat{T}=\hat{\Theta}_1\hat{\Theta}_2$ does both. \footnote{The classification for the superconductors are identical to that for the insulators.} The upper three rows show that the classification reduces to that of spinless fermion if $\hat{\Theta}_2$ symmetry is absent. The lower rows show new possibilities that emerge upon taking spins into account.}
\label{table:spinful}
\end{table*}

Now we are in a good position to consider symmetry changes. 
Important to note here that spontaneously ordered phases can be considered alongside systems under external field, as once a system is deep inside the ordered phase it can be treated within mean-field theory. 

We first consider the symmetry reduction possibilities while maintaining spin degeneracy (see Table \ref{table:spinless}).
 If we only break the time reversal symmetry, the system is characterized by two independent topological indices $(C_K,C_{K'})$, hence the classification is given by $\mathbb{Z}\oplus \mathbb{Z}$.  We refer to these SPT phases as intra-valley quantum (anomalous) Hall (QAH) states. Such phases may be realized by placing trigonally-strained graphene \cite{Levy30072010,Guinea:2010fk} under an external magnetic field, as the sign of the pseudo-magnetic field caused by strain is opposite for the two valleys. Further reducing the symmetry by introducing inter-valley scattering leads to inter-valley QAH state indexed by a single integer $\mathbb{Z}$: the total Chern number~\footnote{As far as symmetry of the phase is concerned, quantum Anomalous Hall state and quantum Hall state are equivalent.}. Breaking electron number conservation turns the above insulators into superconductors. Following the procedure above, we obtain the same classification for the topological superconductors in 2D, resulting in topological valley superconductor (TVSC) and  intra- or inter-valley topological superconductors. Table \ref{table:spinless} summarizes all symmetry reduction possibilities and their classifications starting form gated multi-layer graphene.

Now we consider extending our classification to take the electron spin into account as a dynamical degree of freedom. This will allow us to consider interaction effects at the level of spin ordering.  With both spin and valley degrees of freedom, there are three symmetry operators related to time reversal:   $\hat{\Theta}_1\propto\tau_x$ exchanges two valleys, $\hat{\Theta}_2\propto i\sigma_y$ which acts on the spin indices flips spin, and $\hat{T}=\hat{\Theta}_1\hat{\Theta}_2$ does both. If both $\hat{\Theta}_2$ and $\hat{T}$ are broken, the classification reduces to that of spinless electrons (see upper part of Table \ref{table:spinful}). However, taking any of these symmetry operators into account leads to new classes (see lower part of Table \ref{table:spinful}).

When all $\hat{\Theta}_1$, $\hat{\Theta}_2$ (and as a result $\hat{T}$) and no valley-mixing symmetries are imposed the state corresponds to 
 the so-called ``layer antiferromagnetic'' (LAF) phase predicted in Refs.~\cite{DW-11, PhysRevB.82.205106} and possibly occurring as a ground state in bi-layer graphene at neutrality point~\cite{VelascoJ.:2012uq}. In this phase the product of spin and valley of edge quasi-particles is locked to their momentum. Therefore, one may index this state by its {\em spin-valley Chern number} $C_{SV}$~\cite{DW-11}.  In this phase, for each valley, the edge states associated with two spins are counter-propagating. As long as these two counter-propagating modes do not couple, there can be any number of them per each valley, leading to $2C_{SV}{\rm e^2/h}$ spin-valley Hall conductivity~\cite{DW-11,PhysRevLett.99.236809}.  However there is a form of symmetry allowed coupling between a pair of sets of counter propagating edges; this coupling can gap out the edge modes~\cite{PhysRevLett.95.146802}. Therefore, the number of symmetry protected edge modes for each valley is $C_{SV}$ mod $2$. Hence, we obtain a $\mathbb{Z}_2$ classification for this phase labeled by $(-1)^{C_{SV}}$ as supposed to $\mathbb{Z}$ classification which would be implied by Refs.~\cite{DW-11,PhysRevLett.99.236809}.
 
Another interesting possibility is breaking $\hat{\Theta}_1$, and $\hat{\Theta}_2$, while respecting their product $\hat{T}$ and no-valley mixing symmetry. This leads to the quantum spin Hall (QSH) phase, in which $C_{K,\uparrow}=-C_{K',\downarrow}$, and $C_{K',\uparrow}=-C_{K,\downarrow}$ due to $\hat{T}$ symmetry, while there is no constraint on the $C_{K,\uparrow}+C_{K,\downarrow}$. Hence, unlike usual 2D QSH states~\cite{PhysRevLett.95.226801,PhysRevLett.96.106802}, we obtain $\mathbb{Z}$. {This is because multiple type of time reversal operators can be defined~\cite{PhysRevB.86.205116}} in the presence of no valley mixing symmetry. This and other SPT possibilities are summarized in Table \ref{table:spinful}. 

With the classification at hand, we return to what it implies for the fate of the gapless edge states at the tilt boundary of chirally stacked and gated N-layer graphene. In general, a topological phase transition between two phases within the same class requires the bulk gap to close and reopen (as in inter-plateaux transitions in integer quantum Hall effect). Gapless edge states are guranteed at a physical boundary between such two phases. The edge states at the AB-BA tilt boundary we established in the section II are examples of such edge states. Hence the edge states will remain gapless as long as the time (valley) reversal and charge conservation symmetry are maintained. 

On the other hand, symmetry change can either yield a trivial phase which does not support an edge state or a different type of SPT with different type of edge states. According to Tables I and II, ruining the no valley mixing symmetry is the only way to render the system trivial and gap the edge states. However this requires large momentum transfer which generally requires fine tuning unless the unit-cell becomes enlarged either for the entire system~\cite{Kekule-1,Kekule-2} or for the edge through arm-chair edge, or  short-range disorder such as a vacancy breaks $A$-$B$ sub-lattice symmetry~\cite{Disorder-1}. 
As both rarely occur,  we anticipate gapless edge states at most tilt boundaries. In particular, the natural zigzag boundary formation for the tilt boundaries make such edge states more robust than the edge states in gate polarity boundaries~\cite{Martin_Topological_2008}.

Among symmetry change possibilities leading to another SPT, transition from QVH with spin degeneracy to LAF phase where spin is a dynamic degree of freedom is of particular interest as LAF is suspected to be the ground state of bi-layer graphene near neutrality point~\cite{VelascoJ.:2012uq}. Upon this phase transition the nature of edge states change from spin degenerate valley helical states (QVH) to spin-valley Hall edge states (LAF). In such transitions, the bulk gap has to close and reopen; this is indeed seen in the experiment of~\textcite{VelascoJ.:2012uq}.

\section{Connection to experiments}
\label{sec:expt}
In this section, we discuss the experimental implications of our findings. Specifically we propose the transport through the network of tilt boundaries as a solution to the long standing mystery of sub-gap transport\cite{Oostinga:2008fk,doi:10.1021/nl102459t}. Further we propose feasible experiments to test the proposal.

In order to discuss the topological transport through the network of tilt boundaries observed in Refs.~\cite{McEuen2012,2012arXiv1207.5427H,Brown2012} 
 we should first discuss the effect of the arbitrary angle between the Burger's vector $\vec{b}$ and the tangent vector $\vec{t}$. Microscopic study of tilt boundaries at various angles between $\vec{b}$ and $\vec{t}$ will be presented in the future\cite{liang-pre}.
However, it has been known that a single domain Bernal stacked gapped bilayer ribbon should support gapless edge state for zig-zag edges, but not for arm-chair edges~\cite{PhysRevLett.100.026802,DW-7}.
This is because the arm-chair edge enlarges the unit-cell along the direction parallel to the boundary and makes the projection of  K and K$'$ valley identical. However, \textcite{DW-7} showed that the polarity boundary edge states only develop barely visible gap which is orders of magnitude less then the bulk gap even for a sharp boundary, and the gap decreases quickly when the polarity boundary becomes smooth. 
These arguments apply to our tilt boundary  and  the edge state will develop a small gap when $\vec{b}\parallel\vec{t}$. However, given the large width of the observed tilt boundaries we expect that all straight tilt boundaries will have nearly gapless edge states except those with small angles between $\vec{b}$ and $\vec{t}$. 

When the tilt boundary meanders and changes directions, likely there will be portions with small gap segmenting the gapless regions and the transport will occur through hopping between the gapless regions. The observed 2D network of such tilt boundaries would yield 2D variable range hopping temperature dependence $R(T)\propto \exp(T_0/T)^{1/3}$\cite{VRH-1,VRH-3} at low temperatures governed by the 2D connectivity and the small characteristic gaps of gapped regions. This explains the observed temperature $T$ dependence of resistance at low temperatures~\cite{Oostinga:2008fk,doi:10.1021/nl102459t}.

We propose following experiments to test our proposal. (1) Four terminal transport measurements with two of the contacts, say contacts 1 and 3, at two ends of a tilt boundary. This would yield highly anisotropic transport proving dominant transport along the tilt boundary i.e., $R_{1,3}\ll R_{2,4}$. (2) Scanning tunneling spectroscopy measurements of local density of states. This should measure a gapless spectrum at the tilt boundary but exhibit a gapped spectrum with the gap magnitude of the optical gap away from the tilt boundary.  (3) Thermoelectric imaging. The mid-gap density of state at tilt boundaries would appear in scanning thermopower images. Unpublished thermopower imaging data by \textcite{thermo} indeed show a network with local density of state near fermi energy, that is reminiscent of the tilt boundary network. 
(4) Edge current imaging using scanning SQUID which can detect magnetic field generated by edge currents. 
\section{Conclusion}
\label{sec:summary}
We showed that spin-degenerate tilt boundaries of gated multi-layer graphene support topological gapless edge states protected by three symmetries: time (valley) reversal, no-valley-mixing, and electron number conservation. We demonstrated the existence of gapless edge states through a tight-binding model calculation and a first principal calculation, where the latter took strain effects into account. We then addressed the symmetry protection of the edge states and consequences of symmetry changes within the framework of SPT~\cite{Classification-3}. 

The framework of SPT allowed us to place the 2D topological phase supporting the edge states, namely QVH, among various topological insulator/superconductor phases alongside previously postulated QAH, LAF and QSH. While previous literature postulated QVH, QAH, LAF and QSH to be all supporting number of edge states growing with the number of layers $N$ (i.e. ${\mathbb Z}$-type in the language of classification), we found that the symmetry of  LAF only protects odd number of edge modes for each valley. Hence LAF is a $\mathbb{Z}_2$-topological insulator much like quantum spin Hall insulator\cite{PhysRevLett.95.146802}. Transition between these different SPT's require closing and re-opening of the bulks gap as already been observed in Ref.~\cite{VelascoJ.:2012uq}. 

We predict the naturally occurring tilt boundary~\cite{2012arXiv1207.5427H,Brown2012,McEuen2012} to be the first topological structural defect hosting topologically protected gapless mode of transport, 
Most importantly, our findings on tilt boundaries combined with the recent observations \cite{2012arXiv1207.5427H,Brown2012,McEuen2012} solve the long standing mystery of sub-gap transport~\cite{Oostinga:2008fk}. Our explanation can be tested through proposed transport, scanning tunneling spectroscopy and thermopower imaging experiments, and scanning SQUID experiments. Experimental confirmation of the tilt-boundary transport origin of the sub-gap transport will open doors to control the sub-gap transport and enable device application of gated multi-layer graphene systems. 
 
\vspace{5mm}
{\it Note added.}  After completion of this work, a
complementary preprint\cite{zhang-mele}, which covers material closely related material, has appeared.

\vspace{5mm}
{\bf Acknowledgements}: We thank Paul McEuen for numerous discussions and sharing his unpublished data. We thank Joe Stroscio for sharing his unpublished data on thermopower imaging. We thank E.J. Mele for useful discussions and sharing the preprint \cite{zhang-mele} and J. Sethna for useful discussions regarding structural topological defect aspect of the tilt boundary. 
E.-A.K. and D.N. were supported by NSF Award EEC-0646547 through Cornell Center for Nanoscience. E.-A.K. and A.V. were supported in part by NSF CAREER grant DMR-0955822. E.-A.K. was also supported in part by NSF Grant DMR-1120296 through Cornell Center for Materials Research. Y.L. and L.Y. were supported by NSF Grant No. DMR-1207141. The computational resources have been provided by Lonestar of Teragrid at the Texas Advanced Computing Center (TACC).


\end{document}